\documentclass[aps,twocolumn,showpacs,amsmath,amssymb,prl]{revtex4-1}
\usepackage[english]{babel}
\usepackage{graphicx} 
\usepackage{times}
\usepackage{bbold}
\usepackage{color}
\usepackage{lipsum}

\newcommand{\cl}[1]{\hat{\mathcal{#1}}}

\newcommand{\bra}[1]{{\langle #1 \vert}}

\newcommand{\ket}[1]{{\vert #1 \rangle}}

\newcommand{\upa}[1]{\uparrow_{#1}}
\newcommand{\doa}[1]{\downarrow_{#1}}
\newcommand{\mathbfx}[1]{{\color{red}#1}}
\newcommand{\RR}{\mathbb{R}}
\def\tr{{\,{\rm tr}}}

\begin{document} 
\title{Exact asymptotics of the current in boundary dissipated quantum chains in large external fields}

\author{Zala Lenar\v ci\v c$^{1}$ and Toma\v{z} Prosen$^{2}$}
\affiliation{$^1$J.\ Stefan Institute, SI-1000 Ljubljana, Slovenia}
\affiliation{$^2$Faculty of Mathematics and Physics, University of
Ljubljana, SI-1000 Ljubljana, Slovenia}

\begin{abstract}
Boundary driven quantum master equation for a general inhomogeneous (non-integrable) anisotropic Heisenberg spin $1/2$ chain, or an equivalent nearest neighbor interacting spinless fermion chain, is considered in the presence of a strong external field $f$. We present an exact closed form expression for large $f$ asymptotics of the current in the presence of pure incoherent source and sink dissipation at the boundaries. In application we demonstrate arbitrary large current rectification in the presence of interaction. \end{abstract}


\maketitle

\textit{Introduction.--}
The study of quantum dynamics far from equilibrium is currently at the forefront of theoretical \cite{polkovnikov} (but also experimental \cite{jorg,bloch}) condensed matter physics research.
Although most of non-equilibrium phenomena studied are directly time dependent, such as e.g. in the so-called quantum quenches \cite{calabresecardy}, there is a growing interest in studying steady-state phenomena \cite{essler2014,caux} especially under large thermal/chemical bias and/or external fields \cite{doyon,jaksic,amaricci12,eckstein13}.
A prominent examples of this sort are dielectric breakdown in the Mott insulator \cite{taguchi00,aoki,eckstein,meisner2010,zala1,oka2012,camille12,mazza14} or rectification effects in nanoscale heat or particle transport \cite{casati,baowen,chang}. Particularly fruitful direction which has been developed recently \cite{wichterich07} is a hybrid approach of open quantum systems formalism \cite{petruccione02}, where the bulk of a finite (possibly large) interacting quantum system is treated fully coherently with a {\em unitary} generator, whereas the environment is
considered to act locally on the system's boundaries in terms of incoherent markovian quantum noise.
Remarkably, it has been demonstrated that such an approach allows for interesting exactly solvable, integrable instances \cite{prosen2008,prosen2011b,enej,prosen2014} and can be phenomenologically justified even for non-small system-environment couplings in terms of the so-called {\em repeated interactions} protocol  \cite{karevski}.

An exact treatment of nonequilibrium transport has been so far possible only for {\em pseudo-force} driving originated from the bias in markovian processes at the system's boundaries.
In this Letter we combine this approach with a {\em real force} $f$, resulting from an external field gradient, and formulate a systematic perturbation theory of the steady state current in $1/f$. We show that under a plausible assumption about the zeroth order of the steady state density matrix an explicit asymptotic $f\to\infty$ expression for the  current can be derived for a general inhomogeneous (and nonintegrable) $XXZ$ spin $1/2$ chain. For long chains one can demonstrate arbitrarily large relative current rectification under reversal $f\to -f$. Our formula is clearly corroborated by numerical computations on small chains, where further resonance effects of unusually large currents at particular values of $f$ are discussed and partly explained. The result applies as well for large anisotropy (interaction) asymptotics.

\textit{Quantum transport in boundary \& field driven spin chain.--} 
We consider an open inhomogeneous $XXZ$ spin 1/2 chain exposed to an additional  magnetic field, so that bulk of the system is governed by the Hamiltonian 
$H=H_{xxz} + V$,
\begin{align}
&H_{xxz}= H_{xx} + H_{z},\quad
H_{xx}=\sum_{j=1}^{n-1}2\alpha_j(\sigma_j^+ \sigma_{j+1}^- + \sigma_j^- \sigma_{j+1}^+), \notag\\
& H_z = \sum_{j=1}^{n-1} \Delta_j \sigma_j^z \sigma_{j+1}^z,\quad V=\sum_{j=1}^n f_j \sigma_j^z.
\end{align}
where parameters $\alpha_j,\Delta_j$ may be site dependent. For the {\em homogeneous} case $\alpha_j=1,\Delta_j=\Delta$ will be used. $f_j$ determine the 
profile of magnetic field. Using Wigner-Jordan transformation the problem can be mapped to spinless fermion chain where $\alpha_j$ map to free hopping amplitudes, $\Delta_j$ to interactions and $f_j$ to electric potential.
Density matrix $\rho$ of the system may be described using the Liouville master equation in the Lindblad form \cite{petruccione02,wichterich07}
\begin{align}
&\frac{d\rho(t)}{dt}
=\hat{\mathcal{L}}\rho
:=-i[H,\rho(t)] + \hat{\mathcal{D}}\rho,
\end{align}
where 
$\hat{\mathcal{D}}\rho=\sum_{k} 2 L_{k}\rho(t) L_k^\dagger - \{L_{k}^\dagger L_k,\rho(t)\}$ describes coupling to the bath at the edges.
Here, only the extreme case of pure source/sink on the left/right end, 
$L_1=\sqrt\varepsilon\sigma_1^+, L_2=\sqrt\varepsilon\sigma_n^-$,
will be considered. 
Nonequilibrium steady state (NESS) 
$\rho_\infty=\lim_{t\rightarrow\infty}\rho(t)$
is found as the fixed point $\hat{\mathcal{L}}\rho_{\infty}=0$. 
The effect of driving will be characterized by the spin current from site $j$ to $j+1$
\begin{equation}
J_j=i\alpha_j(\sigma_j^+ \sigma_{j+1}^- - \sigma_j^- \sigma_{j+1}^+),
\end{equation}   
which is in the limit $t\rightarrow \infty$, i.e. for NESS, site independent, $J=\tr(J_j \rho_\infty)/\tr\rho_\infty$. 

\textit{Large field, large interaction asymptotics.--}
We split the Liouvillian into {\em diagonal} and {\em off-diagonal} parts
$\hat{\mathcal{L}} = f \hat{\mathcal{L}}_{0} + \hat{\mathcal{L}}_1 $,
\begin{equation}
\hat{\mathcal{L}}_{0} \rho = -i[\tilde{V} + \tilde{H}_z,\rho], \quad
\hat{\mathcal{L}}_{1} \rho = -i[H_{xx},\rho] + \hat{\mathcal{D}}\rho,  
\end{equation}
where $\tilde{H}_z = \sum_{j=1}^{n-1}\tilde{\Delta}_j \sigma^z_j \sigma^z_{j+1}$,
$\tilde{V}=\sum_{j=1}^n \tilde{f}_j \sigma^z_j$ with $\tilde{\Delta}_j \equiv \Delta_j/f$, $\tilde{f}_j \equiv f_j/f$.
We consider $1/f$ as a formal small parameter, assuming $f\gg \alpha_j,\varepsilon$.
Note that both diagonal generators, $V$ or $H_z$, need not be large. We may either fix $\tilde{f}_n-\tilde{f}_1=1$, in which case $f$ represents the total
field/potential drop across the system while $\Delta_j = {\cal O}(1)$, or consider the large anisotropy (interaction) asymptotics $\tilde{\Delta}_j = {\cal O}(1)$ 
while $\tilde{f}_j={\cal O}(f^{-1})$.
Expanding NESS in powers of $f^{-1}$,
$\rho_{\infty}= \sum_{p=0}^{\infty} f^{-p} \rho^{(p)},$
different orders are connected with the recurrence relation:
\begin{equation}\label{EqLRho0}
\cl{L}_0 \rho^{(p+1)} + \cl{L}_1\rho^{(p)}=0, 
\quad \cl{L}_0 \rho^{(0)}=0.
\end{equation}
Each order of $\rho_{\infty}$ can be expanded in terms of a complete Pauli basis
$
\rho^{(p)}=
\sum_{\{s_1,\dots,s_n\}} M_{s_1,\dots,s_n}^{(p)} \prod_j \sigma^{s_j}_j,
$
where $s_j\in\{0,z,+,-\}$, and $\sigma^0=\mathbb{1}$. 
Condition for $\rho^{(0)}$ in Eq.~(\ref{EqLRho0}) is satisfied for exponentially many products $\prod_j \sigma^{s_j}_j$ for which the sum
$\sum_{j=1}^n f_j\xi_{s_j} +\sum_{j=1}^{n-1} \Delta_j \left((\xi_{s_j}+\xi_{s_{j+1}})\pmod{2}\right) $ vanishes,
where 
$\xi_{s_j} = \pm 1$ for $s_j=\pm$ and $\xi_{s_j}=0$ otherwise. However,  exact numerical diagonalization of $\cl{L}$ on small systems of even size $n\le 8$ shows that for strong field or anisotropy $\rho^{(0)}$ is a diagonal projector
\begin{equation}\label{EqRho0}
\rho^{(0)}= a_0 \prod_{j=1}^{\frac{n}{2}} \frac{1}{2}(\mathbb{1}_j+\sigma_{j}^z) \prod_{j=\frac{n}{2}+1}^{n} \frac{1}{2}(\mathbb{1}_j-\sigma_{j}^z)	
\end{equation}
with step-like magnetization profile. 
In case of large anisotropy this has already been observed \cite{benenti} and given physical motivation suitable for our limit as well.
Thus, we conjecture that $\rho^{(0)}$ has the form of Eq.~(\ref{EqRho0}) for a system of arbitrary even length $n$. If $n$ is odd $\rho^{(0)}$ contains two terms of the form (\ref{EqRho0}) with a jump in magnetization between sites $(\frac{n-1}{2},\frac{n+1}{2})$ or $(\frac{n+1}{2},\frac{n+3}{2})$. Further on we assume even $n$, generalizing the solution to arbitrary $n$ at the end of derivation. 

We introduce the following notation for local operators
\begin{align}\label{EqNotat}
&\frac{1}{2}(\mathbb{1}+\sigma^z)
\equiv
\bigl(
\begin{smallmatrix}
\upa{} \\
\upa{}
\end{smallmatrix} 
\bigr),
\quad 
\sigma^-
\equiv
\bigl(
\begin{smallmatrix}
\doa{} \\
\upa{}
\end{smallmatrix} 
\bigr), \notag \\
&\frac{1}{2}(\mathbb{1}-\sigma^z)
\equiv\bigl(
\begin{smallmatrix}
\doa{} \\
\doa{}
\end{smallmatrix} 
\bigr),
\quad
\sigma^+
\equiv\bigl(
\begin{smallmatrix}
\upa{} \\
\doa{}
\end{smallmatrix} 
\bigr),
\end{align}
i.e., $\bigl(\begin{smallmatrix}\nu \\ \mu\end{smallmatrix}\bigr)\equiv \ket{\nu}\bra{\mu}$, $\nu,\mu\in\{\upa{},\doa{}\}$.
E.g.,  $\rho^{(0)}$ is represented as a single component
\begin{equation}
\rho^{(0)}
=a_0
\bigl(
\begin{smallmatrix}
\upa{}  & \dots & \upa{} \\
\upa{}  & \dots & \upa{} 
\end{smallmatrix} 
\big|
\begin{smallmatrix}
 \doa{} & \dots & \doa{} \\
 \doa{} & \dots & \doa{}
\end{smallmatrix} 
\bigr).
\end{equation}
Vertical bar marks the center of the system. It is instructive to think of the effect that different parts of $\cl{L}$ have on a chosen term $M^{(p)}_{s_1,\ldots,s_n}\prod_j \sigma^{s_j}_j$ in 
such notation.
The action of $\cl{L}_0$ is diagonal, i.e. it does not change the operator product to which it is applied, while unitary part of $\cl{L}_{1}$  causes exchange of neighboring pairs if different, $(\upa{j} \doa{j+1})$ or $(\doa{j} \upa{j+1})$. Dissipator $\cl{D}$ is the only process to relate imaginary and real part of coefficients involved in the recurrence relation and is trivial if both $\mu, \nu$ are aligned with the boundary driving.

Using the recurrence relation~(\ref{EqLRho0}) higher order $\rho^{(p)}$ can be generated from $\rho^{(0)}$. 
E.g. the first order $\rho^{(1)}$ then consists of 
\begin{equation}\label{EqRho1}
\rho^{(1)}=
a_1
\left[
\bigl(
\begin{smallmatrix}
\upa{}  & \dots & \upa{} & \doa{} \\
\upa{}  & \dots & \upa{} & \upa{} 
\end{smallmatrix} 
\big|
\begin{smallmatrix}
 \upa{} & \doa{} & \dots & \doa{} \\
 \doa{} & \doa{} & \dots & \doa{}
\end{smallmatrix} 
\bigr)
+
\bigl(
\begin{smallmatrix}
\upa{}  & \dots & \upa{} & \upa{} \\
\upa{}  & \dots & \upa{} & \doa{} 
\end{smallmatrix} 
\big|
\begin{smallmatrix}
 \doa{} & \doa{} & \dots & \doa{} \\
 \upa{} & \doa{} & \dots & \doa{}
\end{smallmatrix} 
\bigr)
\right].
\end{equation}
However, we will not strive to built the whole NESS density matrix, instead we will relate perturbatively only its component relevant for the NESS spin current. 

\textit{Steady state spin current.--}
We would like to identify the leading order terms in the current $J=\tr(J_j\rho_\infty)/\tr\rho_\infty$. In our diagrammatic notation $\tr$ is a sum of coefficients at configurations with the same upper and lower index at each site. Therefore $\tr\rho_\infty=a_0 + \mathcal{O}(f^{-2})$. Similarly, $\tr J_j\rho_\infty$ is a sum over configurations in $\rho_\infty$ with $\sigma_j^{\pm} \sigma_{j+1}^{\mp}$ and $\mathbb{1}_{l}\pm\sigma^{z}_{l}$ elsewhere. Suppose we calculate current from the operator $J_{n/2}$. We will show that then the leading contribution comes from the components with $\mathbb{1}_l + \sigma^{z}_l$ for $l<n/2$ and $\mathbb{1}_l-\sigma^{z}_l$ for $l>n/2+1$.

We claim that in order to obtain the current from $\tr J_{n/2} \rho_\infty $ up to $\mathcal{O}(f^{-(n+2)})$ only the terms 
$\tilde\rho=\rho^{(0)}+\rho_{1,R}+\rho_{1,L}+\rho_{2,R}+\rho_{2,L}$ 
in the density matrix are relevant. Expression 
{
\begin{align*}
\rho_{1,R}
=\phantom{0.1cm}\\
\frac{a_{1}}{2f} 
\bigl(
\begin{smallmatrix}
\upa{}  & \dots & \upa{} & \doa{} \\
\upa{}  & \dots & \upa{} & \upa{} 
\end{smallmatrix} 
&\big|
\begin{smallmatrix}
\mathbfx{\upa{}} & \doa{} & \doa{} & \dots & \doa{} \\
\doa{} & \doa{} & \doa{} & \dots & \doa{}
\end{smallmatrix} 
\bigr)
+\hspace{0.15cm} \frac{ia_{n}}{f^n} \hspace{0.15cm}
\bigl(
\begin{smallmatrix}
\upa{}  & \dots & \upa{} & \doa{} \\
\upa{}  & \dots & \upa{} & \upa{} 
\end{smallmatrix} 
\big|
\begin{smallmatrix}
\mathbfx{\upa{}} & \doa{} & \doa{} & \dots & \doa{} \\
\doa{} & \doa{} & \doa{} & \dots & \doa{}
\end{smallmatrix} 
\bigr)
\\
2:\ &\Downarrow \hspace{3.6cm} n: \ \Uparrow  \\
+\frac{a_{2}}{f^{2}} 
\bigl(
\begin{smallmatrix}
\upa{}  & \dots & \upa{} & \doa{} \\
\upa{}  & \dots & \upa{} & \upa{} 
\end{smallmatrix} 
&\big|
\begin{smallmatrix}
\doa{} & \mathbfx{\upa{}} & \doa{} & \dots & \doa{} \\
\doa{} & \doa{} & \doa{} & \dots & \doa{}
\end{smallmatrix} 
\bigr)
+\frac{ia_{n-1}}{f^{n-1}} 
\bigl(
\begin{smallmatrix}
\upa{}  & \dots & \upa{} & \doa{} \\
\upa{}  & \dots & \upa{} & \upa{} 
\end{smallmatrix} 
\big|
\begin{smallmatrix}
\doa{} & \mathbfx{\upa{}} & \doa{} & \dots & \doa{} \\
\doa{} & \doa{} & \doa{} & \dots & \doa{}
\end{smallmatrix} 
\bigr)
\\
3:\ &\Downarrow \hspace{3.cm} n-1: \ \Uparrow \\
&\hspace{0.15cm} \vdots \hspace{4.32cm} \vdots  \\
\frac{n}{2}:\ &\Downarrow \hspace{2.9cm} \frac{n}{2}+2: \ \Uparrow  \\
+\frac{a_{\frac{n}{2}}}{f^{\frac{n}{2}}} 
\bigl(
\begin{smallmatrix}
\upa{}  & \dots & \upa{} & \doa{} \\
\upa{}  & \dots & \upa{} & \upa{} 
\end{smallmatrix} 
&\big|
\begin{smallmatrix}
\doa{} & \doa{} & \dots & \doa{} & \mathbfx{\upa{}} \\
\doa{} & \doa{} & \dots & \doa{} & \doa{}
\end{smallmatrix} 
\bigr)
+\frac{ia_{\frac{n}{2}+1}}{f^{\frac{n}{2}+1}} 
\bigl(
\begin{smallmatrix}
\upa{}  & \dots & \upa{} & \doa{} \\
\upa{}  & \dots & \upa{} & \upa{} 
\end{smallmatrix} 
\big|
\begin{smallmatrix}
\doa{} & \doa{} & \dots & \doa{} & \mathbfx{\upa{}} \\
\doa{} & \doa{} & \dots & \doa{} & \doa{}
\end{smallmatrix} 
\bigr)
\\
&\hspace{1.3cm} \mathcal{\hat{D}}: \ \Rightarrow 
\end{align*}%
}%
with $a_p \in \RR$ introduces a sequence of terms connected by the recurrence relation (\ref{EqLRho0}), which takes {\em upper} $\upa{n/2+1}$ in $\rho^{(1)}$ (in {\em red} color) to the right boundary and back to its initial position. After each step coefficient involved is for a $f^{-1}$ higher in expansion, being {\em real} until the action of $\mathcal{\hat{D}}$ and {\em imaginary} after. $\rho_{1,L}$ is constructed symmetrically, taking {\em upper} $\doa{n/2}$ to the left boundary and will be expanded in terms of coefficients $b_p$ with $b_1=a_1$. Other $\rho_{2,R(L)}$ originate from the second term of $\rho^{(1)}$ in Eq.~(\ref{EqRho1}) and correspond to shifting {\em lower} $\upa{n/2+1}$ ($\doa{n/2}$) to the right (left) boundary.

From this construction it is clear that the leading contribution to the current is of order $f^{-n}$, namely
\begin{equation}\label{EqCurrAppl}
J= -\frac{\alpha_{\frac{n}{2}} \ 2(a_{n}+b_n)}{f^n \ a_0} + \mathcal{O}(f^{-(n+2)})
\end{equation}
since real coefficients $a_1,b_1$ get subtracted, while $n$th order coefficient in $\rho_{1,R}$ and $\rho_{2,R}$ (equivalently for $\rho_{1,L}$ and $\rho_{2,L}$) are complex conjugated and sum up yielding a factor $2$ in Eq.~(\ref{EqCurrAppl}). Any other operator product term having $\sigma_{n/2}^{\pm}\sigma_{n/2+1}^{\mp}$ and projectors $\mathbb{1}_i+\sigma^{z}_i$ for some $i>n/2+1$ is of higher order in $f^{-1}$ due to obvious additional steps needed to create it from $\rho^{(0)}$. To calculate the spin current we have to express $a_{n},b_{n}$ with $a_0$. In the procedure each step yields an equation that relates coefficients from the consecutive orders via recurrence (\ref{EqLRho0}). Explicit relations obtained from $\rho_{1,R}$ are:\newline
%
Step 1:
\begin{align} \label{EqStep1}
-2i \alpha_{\frac{n}{2}} & a_{0}+ 
2i\left(\tilde\Delta_{\frac{n}{2}-1} + \tilde\Delta_{\frac{n}{2}+1} 
+ \tilde{f}_{\frac{n}{2}} - \tilde{f}_{\frac{n}{2}+1}\right) a_1=0 \notag \\
&a_{1}=
\frac{a_0 \ \alpha_{\frac{n}{2}}}{\tilde\Delta_{\frac{n}{2}-1} + \tilde\Delta_{\frac{n}{2}+1} 
- \tilde{f}_{\frac{n}{2}+1}+\tilde{f}_{\frac{n}{2}}}
\end{align}
Step $k, \ 1<k<n/2$:
\begin{align}
a_{k}=
 \frac{a_{k-1} \ \alpha_{\frac{n}{2}+k-1}}
{\tilde\Delta_{\frac{n}{2}+k-1} + \tilde\Delta_{\frac{n}{2}+k} + \tilde\Delta_{\frac{n}{2}-1} - \tilde\Delta_{\frac{n}{2}} 
- \tilde{f}_{\frac{n}{2}+k}+\tilde{f}_{\frac{n}{2}}}
\end{align}
Step $n/2$:
\begin{align}
a_{\frac{n}{2}}=
\ \frac{a_{\frac{n}{2}-1} \ \alpha_{n-1}}
{\tilde\Delta_{n-1} + \tilde\Delta_{\frac{n}{2}-1} - \tilde\Delta_{\frac{n}{2}} 
- \tilde{f}_{n}+\tilde{f}_{\frac{n}{2}}}
\end{align}
Step $n/2+1$ (action of dissipator ${\mathcal{\hat D}}$):
\begin{align}
&a_{\frac{n}{2}+1}=
- \frac{a_{\frac{n}{2}}\ \varepsilon}
{2\big[\tilde\Delta_{n-1} + \tilde\Delta_{\frac{n}{2}-1} - \tilde\Delta_{\frac{n}{2}} 
- \tilde{f}_{n}+\tilde{f}_{\frac{n}{2}}\big]}
\end{align}
Step $n/2+1+k, 1\le k<n/2$:
\begin{align}\label{EqStepN}
a_{\frac{n}{2}+1+k}=
\frac{a_{\frac{n}{2}+k} \ \alpha_{n-k}}
{\tilde\Delta_{n-k-1} + \tilde\Delta_{n-k} + \tilde\Delta_{\frac{n}{2}-1} - \tilde\Delta_{\frac{n}{2}} 
- \tilde{f}_{n-k}+\tilde{f}_{\frac{n}{2}}}
\end{align}
Combining Eqs.~(\ref{EqStep1} - \ref{EqStepN}) we finally obtain $a_{n}$ as a function of $a_0$ and parameters of Hamiltonian. $b_{n}$ is expressed analogously,
\begin{widetext}
\begin{align}
&
\resizebox{0.90\hsize}{!}{$
\frac{a_n}{a_0}=
- 
\frac{\varepsilon\alpha_{n-1}}{2\left(
\tilde\Delta_{n-1} + \tilde\Delta_{\frac{n}{2}-1} - \tilde\Delta_{\frac{n}{2}} 
- \tilde{f}_{n}+\tilde f_{\frac{n}{2}}\right)^2}
\prod\limits_{k=1}^{\frac{n}{2}-1}
\frac{\alpha_{\frac{n}{2}+k-1} \ \alpha_{\frac{n}{2}+k}}
{\left(
\tilde\Delta_{\frac{n}{2}+k} + \tilde\Delta_{\frac{n}{2}+k-1} 
+\tilde{\Delta}_{\frac{n}{2}-1} - \tilde\Delta_{\frac{n}{2}}
- \tilde{f}_{\frac{n}{2}+k}+\tilde{f}_{\frac{n}{2}}\right)^2}
$}, \label{Equ1i}\\
&
\resizebox{0.90\hsize}{!}{$
\frac{b_n}{a_0}=
-\frac{\varepsilon\alpha_{1}}{2\left(
\tilde\Delta_{1} + \tilde\Delta_{\frac{n}{2}+1} - \tilde\Delta_{\frac{n}{2}} 
- \tilde{f}_{\frac{n}{2}+1}+\tilde{f}_{1}\right)^2}
\prod\limits_{k=1}^{\frac{n}{2}-1}
\frac{\alpha_{\frac{n}{2}-k+1} \ \alpha_{\frac{n}{2}-k}}
{\left(
\tilde\Delta_{\frac{n}{2}-k} + \tilde\Delta_{\frac{n}{2}-k+1} 
+\tilde\Delta_{\frac{n}{2}+1} - \tilde\Delta_{\frac{n}{2}}
- \tilde{f}_{\frac{n}{2}+1}+\tilde{f}_{\frac{n}{2}-k+1}\right)^2} 
$}. \label{Eqd1i}
\end{align}
\end{widetext}
For a homogeneous chain and field which is odd with respect to the center $f_j=-f_{n+1-j}$ expression for the current simplifies
\begin{align}\label{EqJHom}
J
=\frac{2\varepsilon}{f^{n}} \ 
\left(
\tilde\Delta 
- \tilde{f}_{n}+\tilde{f}_{\frac{n}{2}}\right)^{-2}
\prod_{k=1}^{\frac{n}{2}-1}
\left(
2\tilde\Delta
- \tilde{f}_{\frac{n}{2}+k}+\tilde{f}_{\frac{n}{2}}\right)^{-2}.
\end{align}
One should note that a shift of $\tilde{f}_j$ for a constant does not change the result. Obtained expression for the current is appropriate for a general monotonic magnetic field, however, in the case of linear field profile $f_j-f_{j-1} \equiv g$ it has especially simple form in terms of a gradient $g$
\begin{equation}\label{EqJLin}
J=\frac{8\varepsilon}{g^{n}} \ \left(\frac{2\Delta}{g}-n\right)^{-2} \ 
\prod_{k=1}^{\frac{n}{2}-1} \left(\frac{2\Delta}{g}-k\right)^{-2}.
\end{equation}
For systems of odd size $n$ one should consider $J_{(n-1)/2}$ (or $J_{(n+1)/2}$). The leading order contributions originate from the part of $\rho^{(0)}$ with jump in magnetization between sites $(\frac{n-1}{2},\frac{n+1}{2})$ (or $(\frac{n+1}{2},\frac{n+3}{2})$). The imaginary coefficient relevant for the current is then obtained as in system of even size $n-1$, having contribution only from the boundary that is closer to the magnetization jump. The current for odd $n$, homogeneous $XXZ$ chain and field of odd symmetry w.r.t. the center of the system is then simply $J(n)=J(n-1)/4$. For an inhomogeneous chain or non-symmetric monotonic field the two components in $\rho^{(0)}$ have different weights, causing an additional prefactor in the latter equation.
\begin{figure}[ht]
\includegraphics[width=0.4\textwidth]{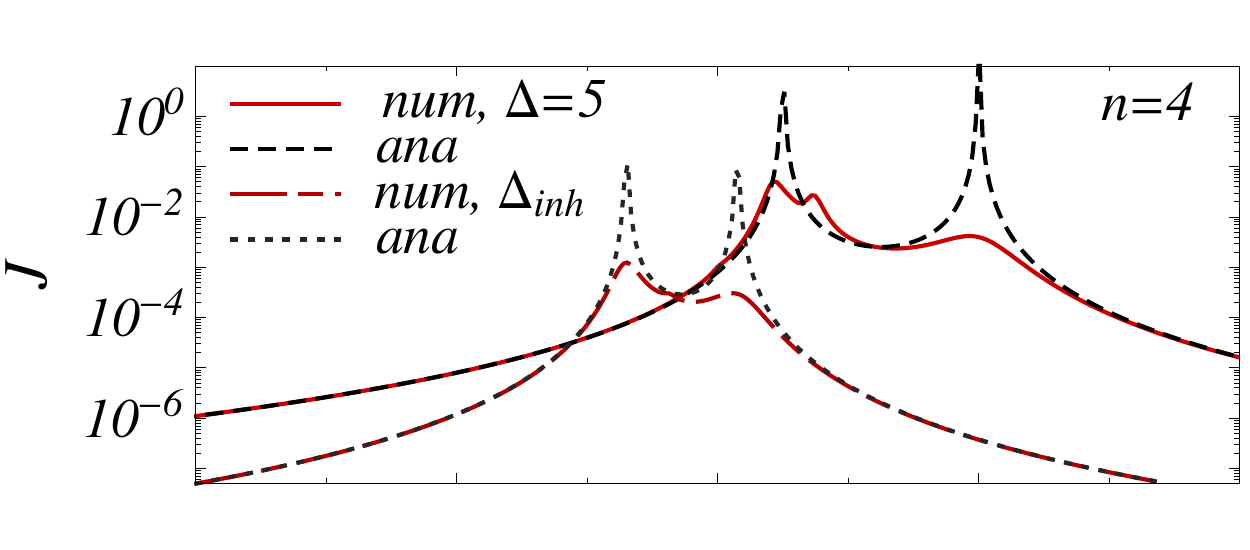}\\\vspace{-0.38cm}
\includegraphics[width=0.4\textwidth]{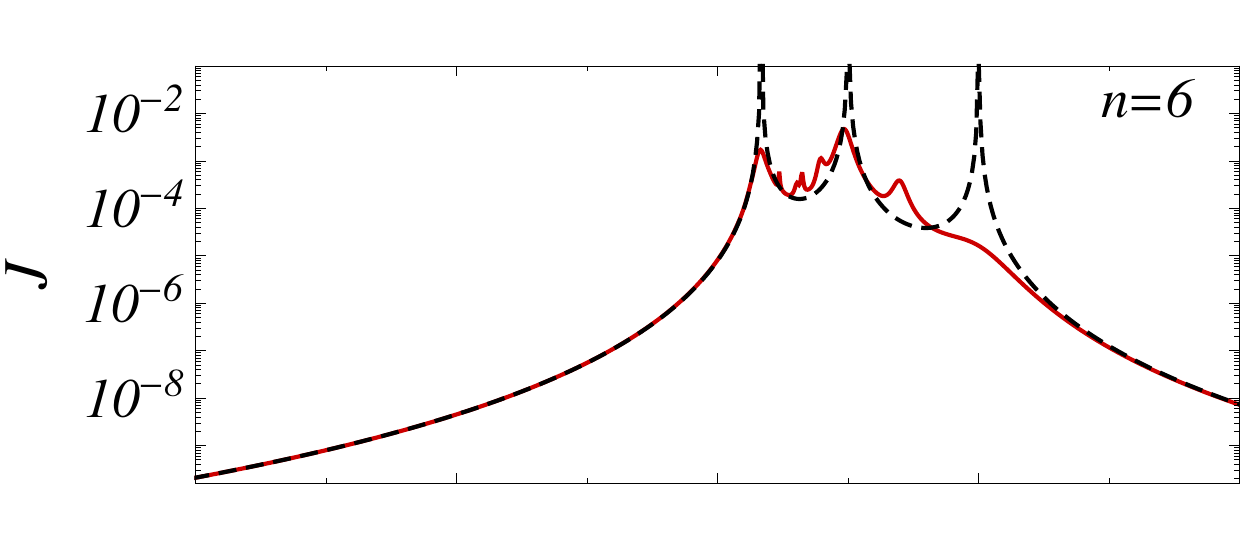}\\\vspace{-0.38cm}
\includegraphics[width=0.415\textwidth]{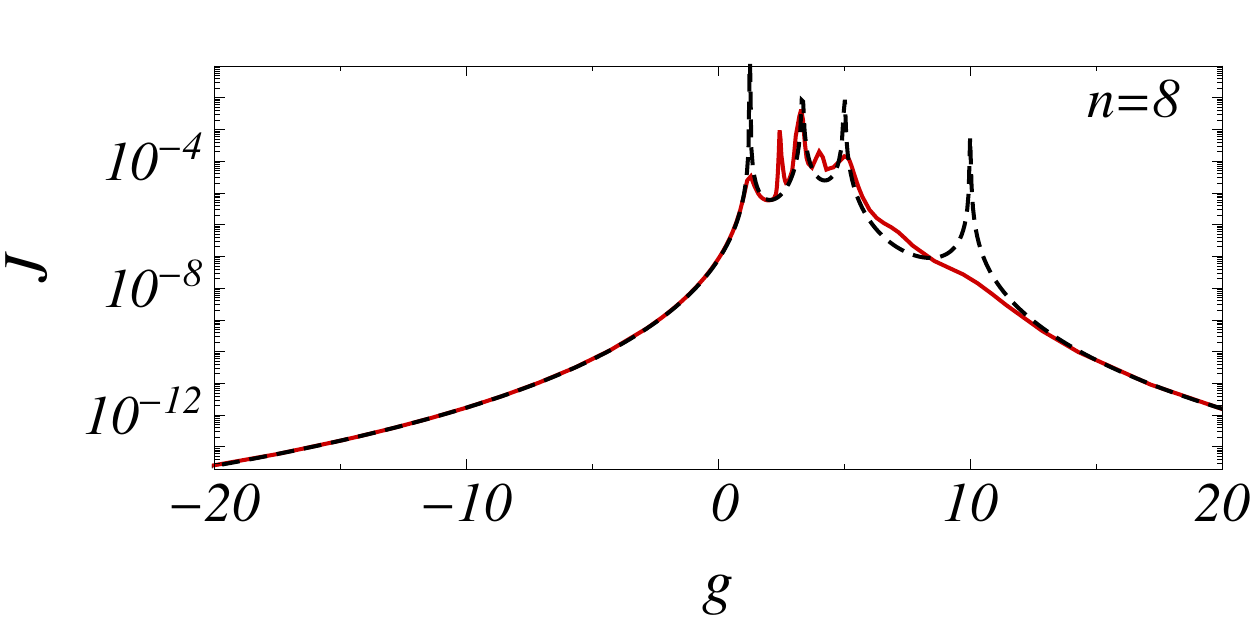}
\caption{(Color online) 
Spin current $J$ as a function of gradient $g$ for linear field profile and anisotropy $\Delta=5$ and $\varepsilon=1$ in systems of size $n=4,6,8$ obtained numerically (num) and from Eq.~(\ref{EqJLin}) (ana). Curve at $n=4$ labeled by $\Delta_{inh}$ shows same comparison for an inhomogeneous 
chain with parameters given in the text.
}
\label{Fig2}
\end{figure}
\begin{figure}[ht]
\includegraphics[width=0.4\textwidth]{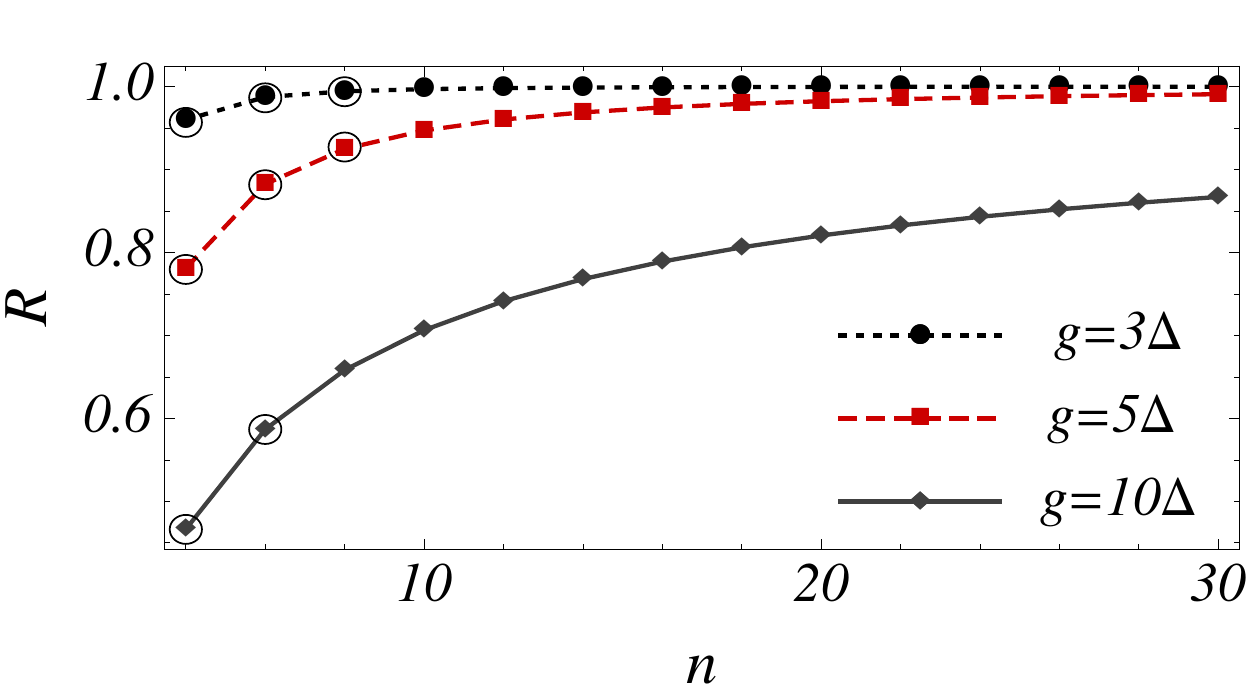}
\caption{(Color online) Dependence of flux rectification $R$ on system size $n$ for $\Delta>0,\varepsilon=1$ and gradients $g=3\Delta,5\Delta,10\Delta$ obtained from Eqs.~(\ref{EqRectDef},\ref{EqJLin}). Circles correspond to exact numerics for small systems.}
\label{Fig3}
\end{figure}

In Fig.~\ref{Fig2} we compare numerical and analytical results for the spin current $J$ at linear field profile and homogeneous Hamiltonian with $\Delta=5$ in systems of size $n=4,6,8$. At $n=4$ we show also an example with inhomogeneous Hamiltonian with parameters 
$\{\alpha_1,\alpha_2,\alpha_3\}=\{0.1,0.3,0.5\}, \{\Delta_1,\Delta_2,\Delta_3\}=\{-2.5,-5,-1\}$. Negative $\Delta_j$ are chosen for clarity since 
$J(-\Delta_j,-f_j)=J(\Delta_j,f_j)$.
Evidently, expression~(\ref{EqJLin}) perfectly captures the dependence $J(g)$ at large enough gradients $g$ and indicates the position of some of the resonances in the numerical result. 
From the construction of the perturbative solution it is clear that poles in Eqs.~(\ref{Equ1i},\ref{Eqd1i}) appear when at some level of perturbative hierarchy the action of diagonal $\cl{L}_0$ on considered operator product is trivial. Consequently, the hierarchy is disconnected and our expression not valid any more. 
Additional resonances observed in numerical results originate from other operator products which are annulled by $\cl{L}_0$ that are not contained in $\tilde\rho$.
In the resonances the hierarchy can be altered in a way that yields a contribution to the current in lower order of $f^{-1}$ \cite{resonance}.
The range of validity of asymptotic expression (\ref{EqJLin}), which is $g \in(-\infty,0) \cup (g_c,\infty)$ for $\Delta>0$, is set by the furthest resonance, yielding a critical gradient 
$g_{c}=2\Delta$.

\textit{Flux rectification.--}
Asymmetry of the driving is revealed by changing the direction of the driving force, $f \rightarrow -f$, resulting in 
$J(f)\neq J(-f)$.
Rectifying behavior is conveniently quantified by rectification coefficient \cite{landi14}
\begin{equation}\label{EqRectDef}
R=\frac{J(f)-J(-f)}{J(f)+J(-f)},	
\end{equation}
which is zero in case of no rectification and reaches extreme values $\pm 1$ when the system for fields in one direction behaves as a perfect insulator. 

For one direction of the dissipative Lindblad driving the anisotropy $\Delta_j$ and the field $f_j$ have a cooperative effect and competing one for the other direction \cite{resonance2}. Fixing dissipative driving to our initial choice and assuming $\Delta_j <0$, when the interaction prefers aligned spins, magnetic field with $g>0$ dictates the same orientation of spins as the Lindblad driving, resulting in an enhanced insulating behavior of NESS. If $g<0$ their effects are contrary, manifested in the enlarged spin current. Relation $J(-\Delta_j,-f_j)=J(\Delta_j,f_j)$ generalizes the argument to $\Delta_j>0$. 

For gradients above the critical value the rectification coefficient $R$ can be calculated for any system size $n$ using the current asymptotics (\ref{Equ1i}-\ref{EqJLin}). In Fig.~\ref{Fig3} we plot the dependence of $R$ on the system size $n$ (even only) for a homogeneous chain and linear field profile at three different gradients, $g=3\Delta,5\Delta,10\Delta$. Circles correspond to exact results for small systems.
Our results clearly show that in the thermodynamic limit $R$ reaches the extreme value
$\lim_{n\rightarrow \infty}R=\textrm{sgn}(\Delta)$,
corresponding to insulating behavior for gradients of opposite sign than anisotropy $\Delta$.

\textit{Discussion.--}
We have developed a systematic asymptotic expansion of the spin/particle current in quantum spin/fermion chains with nearest neighbor interaction driven with inhomogeneous external fields and dissipated with incoherent source/sink baths at the chain ends.
Our results have been worked out in detail on an example of an inhomogeneous $XXZ$ spin 1/2 chain, driven by a transverse magnetic field gradient.
We can point out two mechanisms for enhancing current rectification: (i) rectification coefficient $R\to 1$ in thermodynamic limit $n\to\infty$, at fixed large gradient $f/(n-1)$, for any nonzero anisotropy/interaction $\Delta \neq 0$, as a result of our asymptotic current formula; (ii) for {\em resonant|} field (gradient) values where our asymptotic analysis fails, the current is of higher order in $1/f$ for one direction of the field $f$ than in the other.
This clearly yields rectification coefficient $R\to 1$ even at fixed finite size $n$. Note that there is no rectification effect in noninteracting chains $\Delta=0$, in agreement with~\cite{landi14}.
Our analysis also applies to reveal large anisotropy asymptotics at vanishing external fields by setting $\Delta_j = f \tilde{\Delta}_j, f_j = 0$ in Eqs.~(\ref{Equ1i},\ref{Eqd1i}), yielding for the homogeneous $XXZ$ chain 
$J=\varepsilon \ 2^{-\gamma}\Delta^{-n},\ \gamma=n^2/4-n/2-1$
for even $n$ and 
$J=\varepsilon \ 2^{-\delta}\Delta^{-(n-1)},\ \delta=(n-1)^2/4-(n-1)/2+1$ for odd $n$, 
as noted numerically in \cite{benenti}.

\acknowledgments
Discussions with M. \v Znidari\v c and E. Ilievski, as well as support by the grants P1-0044, J1-5349 of the Slovenian Research Agency are acknowledged.


\begin{thebibliography}{10}
\bibitem{resonance}
Reordering within $\rho_{\infty}$ can be motivated by degeneracy of eigenstates
  of Hamiltonian $H$. It turns out that a resonance in the current appears
  whenever
  $\ket{0}=\ket{\upa{1}\cdots\upa{\frac{n}{2}}\doa{\frac{n}{2}+1}\cdots\doa{n}}$,
  which is nearly an eigenstate for large anisotropy or large field, is
  degenerate with another eigenstate $\ket{i}$ (with less of insulating nature)
  with boundary spins at least partially compatible with the boundary driving.
  Such a degeneracy reduces the insulating effect of $\ket{0}\bra{0}$ and
  promotes $\ket{0}\bra{i},\ket{i}\bra{0},\ket{i}\bra{i}$ in $\rho_{\infty}$.
\bibitem{resonance2}
If the field and the boundary driving try to align spins in the same manner for
  $\Delta<0$ the insulating state with step-like magnetization (compatible with
  source/sink) will not be degenerate with any other eigenstate of $H$ at
  arbitrary field strength \cite{resonance}. In this sense the field and the
  interaction cooperate and separate its energy from other eigenstates.
  By reversing the field or the boundary driving degeneracies become possible. The
  same effect at $\Delta>0$ is obtained with the field of opposite sign.
\end{thebibliography}
\end{document}